\documentclass[sigconf]{acmart}

\usepackage{booktabs} 
\usepackage{subcaption}
\usepackage{multirow}
\usepackage{balance}
\usepackage[para]{footmisc}

\setcopyright{rightsretained}
\newcommand{\OurModel}{AVLEM}
\copyrightyear{2019} 
\acmYear{2019} 
\acmConference[CIKM '19]{The 28th ACM International Conference on Information and Knowledge Management}{November 3--7, 2019}{Beijing, China}
\acmBooktitle{The 28th ACM International Conference on Information and Knowledge Management (CIKM '19), November 3--7, 2019, Beijing, China}
\acmPrice{15.00}
\acmDOI{10.1145/3357384.3357939}
\acmISBN{978-1-4503-6976-3/19/11}

\begin{document}
	\fancyhead{}
	\title{Conversational Product Search Based on Negative Feedback}

\author{Keping Bi$^1$, Qingyao Ai$^1$,  Yongfeng Zhang$^2$, W. Bruce Croft$^1$}
\affiliation{%
	\institution{$^1$College of Information and Computer Sciences, University of Massachusetts Amherst, Amherst, MA, USA}
}
\email{{kbi, aiqy, croft}@cs.umass.edu}
\affiliation{%
	\institution{$^2$Department of Computer Science, Rutgers University, Piscataway, NJ, USA}
}
\email{yongfeng.zhang@rutgers.edu}

	\begin{abstract}
		Intelligent assistants change the way people interact with computers and make it possible for people to search for products through conversations when they have purchase needs. During the interactions, the system could ask questions on certain aspects of the ideal products to clarify the users' needs. For example, previous work proposed to ask users the exact characteristics of their ideal items \cite{zhang2018towards, Sun:2018:CRS:3209978.3210002} before showing results. However, users may not have clear ideas about what an ideal item looks like, especially when they have not seen any item. So it is more feasible to facilitate the conversational search by showing example items and asking for feedback instead. 
		In addition, when the users provide negative feedback for the presented items, it is easier to collect their detailed feedback on certain properties (aspect-value pairs) of the non-relevant items. By breaking down the item-level negative feedback to fine-grained feedback on aspect-value pairs, more information is available to help clarify users' intents. So in this paper, we propose a conversational paradigm for product search driven by non-relevant items, based on which fine-grained feedback is collected and utilized to show better results in the next iteration. We then propose an aspect-value likelihood model to incorporate both positive and negative feedback on fine-grained aspect-value pairs of the non-relevant items. Experimental results show that our model is significantly better than state-of-the-art product search baselines without using feedback and those baselines using item-level negative feedback. 
	\end{abstract}
	
	%
	%


	\maketitle
	
\section{Introduction}
\label{sec:introduction}
People search and browse the products in E-commerce platforms such as Amazon when they have purchase needs. Conventional product search engines return items to users according to their initial queries and do not dynamically interact with users to learn more about their preferences. As a result, users need to browse many products, rewrite queries to specify their needs, or use filters on facets to narrow down results. 
Intelligent assistants such as Google Now, Apple Siri or Microsoft Cortana provide new interaction modes between human and systems, i.e., through conversations. In this way, it becomes possible for an intelligent shopping assistant to actively interact with users to clarify their intents, dynamically refine the ranking, and guide them to find the items they like. An effective intelligent shopping assistant will improve users' search experience substantially and save users much effort spent browsing and filtering to find the ideal items. Thus, in this paper, we focus on the essential part of building an intelligent shopping assistant, constructing an effective conversational product search system. 

To clarify users' shopping intents during interactions, the search system could explicitly ask the users what characteristics they would like the items to have, as proposed in \citet{zhang2018towards,Sun:2018:CRS:3209978.3210002}. For example, when a user expresses the purchase need ``a mobile phone'', the assistant asks the user what brand she likes or  what kind of screen she prefers. With the collected user responses on some aspects of the items, the assistant knows the explicit preferences of the user and refines the ranking to promote relevant items to the top. 

\begin{figure}
	\includegraphics[width=0.4\textwidth]{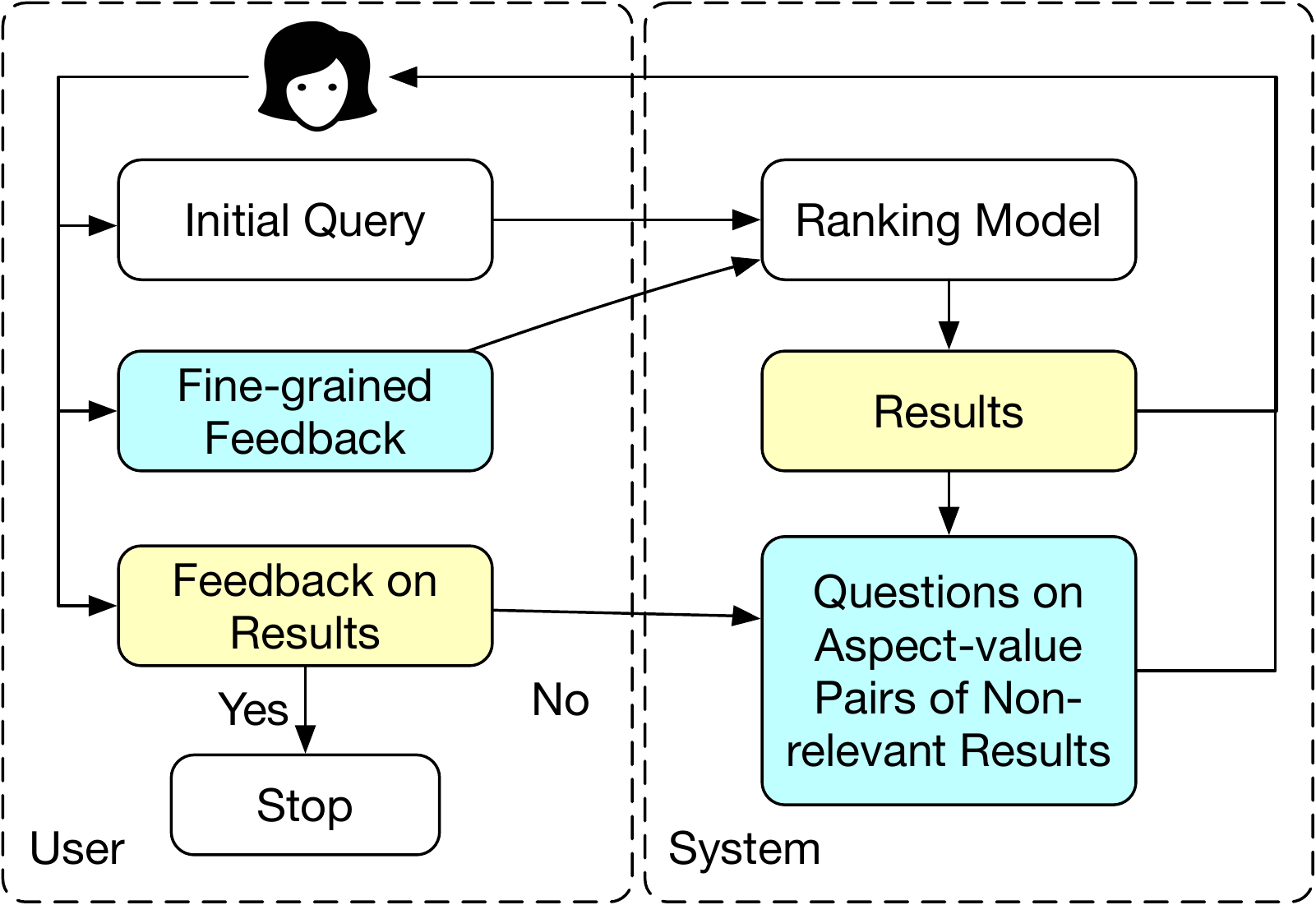} %
	\caption{A workflow of conversational search system based on negative feedback.}
	\label{fig:workflow}
\end{figure}

However, previous work has limitations since users do not always know their exact ideal products when they are shopping, especially before they have seen some examples. 
In contrast, when they are shown an item, they usually know whether they like the item or not. If the item is not good for them, they can tell which aspects they are not satisfied with. 
For instance, when a user who aims to find a mobile phone but do not have preferences on the brand, it is easier for her to answer ``Do you want a curved screen?'' after showing her a phone with curved screen than ``Which kind of screen do you like?'' at the very beginning. So we propose a new paradigm for conversational product search motivated by negative feedback. To be specific, after the user's initial request, several items are shown to the user. If she is not satisfied with the items, her detailed preferences on aspect-value pairs (such as ``brand-Samsung'', ``screen-curved'' and``battery-removable'') of the items are gathered. Then based on the fine-grained feedback on the non-relevant results, the remaining items are re-ranked in the next iteration. This process proceeds until the user finally find the ``right'' product (shown in Figure \ref{fig:workflow}).

\label{sec:paradigm}

Compared with positive feedback, negative feedback is more challenging since relevant results usually have similar characteristics while the reason for a result to be non-relevant could be varied. Previous work on negative feedback \cite{karimzadehgan2011improving,wang2007improve,wang2008study} mainly focuses on document retrieval. They extract negative topic models from the non-relevant documents and demote the results with high similarities to the negative topic models during re-ranking. However, result-level negative feedback is not very informative especially when there are only a few non-relevant results available. In contrast, by collecting fine-grained positive or negative feedback on aspect-value pairs of the non-relevant results, more information can be used as the basis for re-ranking and lead to better performance. Thus, the primary focus of this paper is to effectively incorporate feedback on aspect-value pairs with the ranking model.

Inspired by the idea of the query likelihood model \cite{ponte1998language} which predicts the probability of a query given the document, we propose an aspect-value likelihood model for negative feedback, which predicts the probability of an aspect-value pair being positive or negative and combines it with the original ranking model without using feedback. Specifically, the aspect-value generation model is decomposed to an aspect generation model given an item and a value generation model given the item and aspect. 
Then a multivariate Bernoulli (MB) distribution is assumed for the aspect generation model, and two independent MB distributions are assumed to model the probabilities of positive and negative values respectively given the associated items and aspects. In this way, the negative feedback on an aspect-value pair can be incorporated similarly to positive feedback with a second group of embeddings learned for its multivariate Bernoulli model. 
Our model shows significantly better performance compared with baselines of both item-level negative feedback methods and state-of-the-art neural ranking models for product search without using feedback. We also show the effectiveness of each part of our model through ablation study in \ref{subsec:ablation}. 

\section{Related Work}
\label{sec:related_work}
Three lines of research are related to our work: conversational search and recommendation, product search, and negative feedback in information retrieval (IR).

\textbf{Conversational Search \& Recommendation.} 
\label{subsec:conv_search}
The concepts of conversational search were proposed in some earliest work in IR. \citet{croft1987i3r} designed an intelligent intermediary for information retrieval, named as I$^3$R, which communicates with users during a search session and reacts based on the goals stated by users and their evaluation of the system output. \citet{belkin1995cases} built an interactive IR system, MERIT, that used script-based information-seeking dialogues as interaction for effective search. 

With the emerging of various intelligent conversational assistants in recent years, task-based conversations based on natural dialogues have drawn much attention. 
Radlinski et al. \cite{Radlinski:2017:TFC:3020165.3020183} proposed a theoretical framework with some basic philosophies for conversational IR. Kenter et al. \cite{kenter2017attentive} considered building the representation of conversations as the process of machine reading, based on which answers are retrieved. 
Information-seeking conversations have been collected in \cite{thomas2017misc, qu2018analyzing} and user studies on the collected conversations are conducted to inform the design of a conversational search system \cite{thomas2017misc, qu2018analyzing, trippas2018informing}. 
Mcginity et al. \cite{mcginty2006adaptive} leveraged preference and rating based feedback in a conversational recommender system and emphasize product diversity rather than similarity to conduct effective recommendation. \citet{christakopoulou2016towards} developed a framework to identify which questions to ask in order to quickly learn user preferences and refine the recommendations during the conversations. 
\citet{zhang2018towards} proposed a paradigm for conversational product search, where the system asks users their preferred values of an aspect, shows results when it is confident, and adopts a memory network to ask questions and retrieving results.  
Sun et al. \cite{Sun:2018:CRS:3209978.3210002} proposed a recommendation system based on a similar paradigm, which also collects users' preferred values for given aspects and uses a reinforcement learning framework to choose actions from asking for the values or making recommendations by optimizing a per-session utility function. 

Our research is different from previous work in that $1)$ instead of retrieving answers, we focus on product-seeking conversations; $2)$ in the work where the system also asks users their preferences, they either ask for result-level preference or the explicit values the users prefer for an aspect. In contrast, our system just asks for the users' fine-grained relevance feedback on a given aspect-value pair, which is much easier for users to answer.  

\textbf{Product Search.}
\label{subsec:product_search}
Compared with text retrieval, product search has different characteristics, e.g., product information is more structured and user purchases can be used as labeled data for training and testing. Considerable work has been done based on facets such as brands and categories \cite{lim2010multi, vandic2013facet}. However, free-form user queries are difficult to structure. To support search based on keyword queries, \citet{duan2013probabilistic, duan2013supporting} extended the query likelihood \cite{ponte1998language} method by assuming that queries are generated from a mixture of two language models, one of the background corpus, the other of products conditioned on their specifications. 
This approach still cannot solve the vocabulary mismatch problem between user queries and product descriptions or reviews. \citet{van2016learning} introduce a latent vector space model to alleviate this problem, which learns the vectors of words and products by predicting the products with n-grams in their descriptions and reviews and then matches queries and products in the semantic space. Later, \citet{ai2017learning} noticed that product search can be personalized and proposed a hierarchical embedding model based on product reviews for personalized product search. Recently, \citet{bi2019study} studied different context dependencies in multi-page product search. 

There is also research on other factors such as visual preferences, diversity, and labels for training in product search. \citet{di2014relevance} and \citet{guo2018multi} showed the effectiveness of utilizing images for product search. \citet{parikh2011beyond, yu2014latent} tried to improve product diversity in order to satisfy different user intents behind the same query.  
\citet{wu2018turning} jointly modeled clicks and purchases in a learning-to-rank framework in order to optimize the gross merchandise volume. \citet{karmaker2017application} compared the performance of leveraging click-rate, add-to-cart ratios, order rates as labels for training.

Most work in this line treats product search as a static process, where one-shot ranking is performed based on a user query. In contrast, we focus on a dynamic ranking problem where ranking in the next iteration of the conversation will be refined based on users' fine-grained feedback on the non-relevant items. 

\textbf{Negative Feedback.}
Next, we review methods that retrieve relevant results based only on known non-relevant ones. It is not our focus to use non-relevant results as a complement to relevant ones for identifying extra relevant results.
Previous studies on negative feedback alone mainly focused on document retrieval for difficult queries. 
\citet{wang2007improve} proposed to extract a negative topic model from non-relevant documents by assuming that they are generated from the mixture of the topic model of the background corpus and the negative topic model. Rocchio \cite{rocchio1971relevance} is a feedback method that considers both positive and negative feedback in the framework of the vector space model. It can also be used in the scenarios where only negative feedback is available. \citet{wang2008study} studied negative feedback methods based on the language model and vector space model. Later, \citet{karimzadehgan2011improving} further improved the performance of negative feedback by building a more general negative topic model. 
\citet{peltonen2017negative} introduced a novel search interface, where keyword features of the non-relevant results are provided to users, and they are asked for feedback on the keywords. Then a probabilistic user intent model is estimated to refine re-ranking. 
In addition, \citet{zagheli2017negative} also proposed a language model based method to avoid suggesting results similar to the document users dislike for text recommendation. 

Most previous work on negative feedback only uses result-level non-relevant information except \cite{peltonen2017negative}, which further acquires keyword-level feedback on non-relevant results. Although we also ask users for feedback on finer-grained information, we leverage aspect-value pairs of non-relevant results and focus on product search.


\section{Aspect-value Likelihood Embedding Model for Negative Feedback}
\label{sec:avhem}
\begin{figure}
	\centering
	\includegraphics[width=0.45\textwidth]{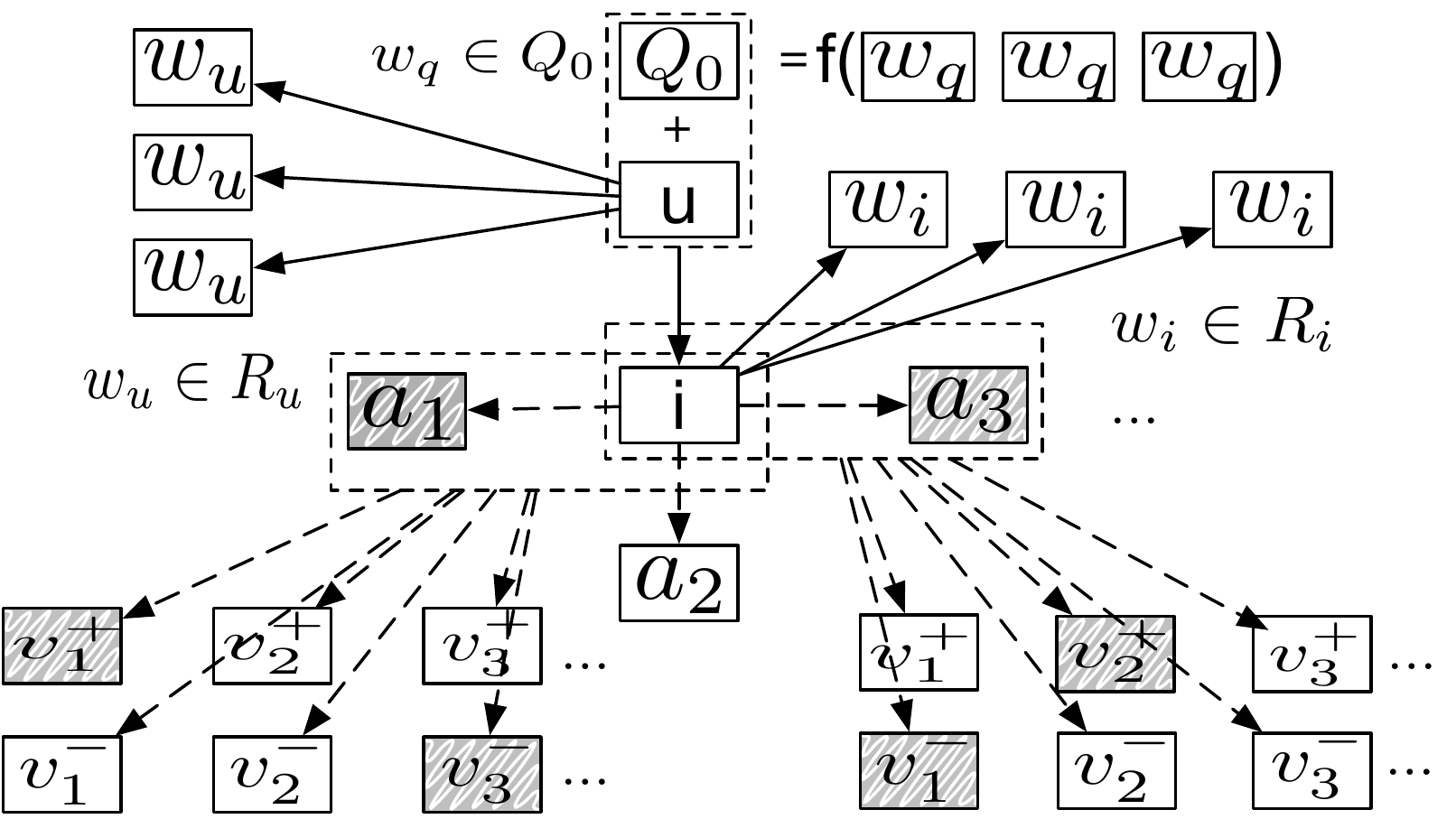} 
	\caption{Our aspect-value likelihood embedding model (\OurModel). The solid and dotted arrows represent the generation from a multinomial and a multivariate Bernoulli distribution respectively. The shaded and blank background represent occurrence and nonoccurrence of the target. $v^+$ and $v^-$ denote positive and negative values. }
	\label{fig:model}
\end{figure}

There are two major modules in our system to conduct product search through conversations with users: selecting aspect-value pairs to ask for feedback and ranking based on the fine-grained feedback. For the aspect-value pair selection, we adopt heuristic strategies, i.e., selecting several random pairs, or pairs mentioned most in the reviews of the non-relevant items, and leave the investigation of other potentially better methods as future work. Then we focus on the ranking model that leverages feedback on aspect-value pairs. We propose an aspect-value likelihood embedding model (\OurModel) which can rank items both with and without feedback. The overall structure of \OurModel~ is shown in Figure \ref{fig:model}. 

We introduce the problem formalization for our task in \ref{subsec:problem_form} and the components of our model in the following subsections.  

\subsection{Problem Formalization}
\label{subsec:problem_form}
A conversation is initiated with a query $Q_0$ issued by a user $u$. In the $k$-th iteration, a batch of results $D_k$ are retrieved and shown to the user. When $D_k$ does not satisfy the user need, from all the shown non-relevant results, $D_1 \cup D_2 \cdots \cup D_{k}$, denoted as $D_{1:k}$, the system extracts a set of aspect-value pairs, namely, $AV(D_{1:k})$.
Then the system selects $m$ aspect-value pairs $\{(a_{k,j}, v_{k,j})|1\leq j \leq m\}$ from $AV(D_{1:k})$ and asks $m$ corresponding questions $\{\mathcal{Q}(a_{k,j}, v_{k,j})|1\leq j \leq m\}$ to the user about whether she likes the aspect-value pairs of the non-relevant results.
After collecting the user's feedback to $\mathcal{Q}(a_{k,j}, v_{k,j})$, denoted as $I(a_{k,j}, v_{k,j})$, in the $k\!+\!1$-th iteration, the goal of the system is to show a list of results $D_{k+1}$, which ranks the finally purchased item $i$ 
on the top. 
The sequence of actions in the conversation can be represented with 
\begin{equation*}
\label{eq:conv_seq}
\begin{aligned}
u \rightarrow Q_0; \; & D_1, \mathcal{Q}_{1,1}, I_{1,1}, \cdots, \mathcal{Q}_{1,m}, I_{1,m}; \; \cdots; \\ 
&D_k, \mathcal{Q}_{k,1}, I _{k,1}, \cdots, \mathcal{Q}_{k,m}, I_{k,m} \rightarrow i
\end{aligned}
\end{equation*}
where $\mathcal{Q}_{k,j}$ and $I_{k,j}$ denote $\mathcal{Q}(a_{k,j}, v_{k,j})$ and $I(a_{k,j}, v_{k,j})$ respectively. $\mathcal{Q}_{k,j}$ is a yes-no question and $I_{k,j}$ can be $1$ or $-1$ to indicate that the answer is yes or no to the question. In addition, reviews of $u$ and $i$ are available to facilitate the ranking, denoted as $R_u$ and $R_i$ respectively.

In this paper, we focus on the scenario where only one result is retrieved during each iteration, namely $|D_k| = 1$. However, the method we propose can cope with general cases with more than one result retrieved in each iteration.


\subsection{Item Generation Model}
We construct an item generation model to capture the purchase relationship between items and their associated users and queries. 
Similar to \cite{ai2017learning}, an item $i$ is generated from a user $u$ and her initial request query $Q_0$. The probability can be computed with the softmax function on their embeddings:
\begin{equation}
\label{eq:item_gen}
P(i|u,Q_0) = \frac{\exp \Big(\mathbf{i} \cdot \big(\lambda \mathbf{Q_0} + (1-\lambda) \mathbf{u} \big)\Big)}
{\sum_{i' \in S_i} \exp \Big(\mathbf{i'} \cdot \big(\lambda \mathbf{Q_0} + (1-\lambda) \mathbf{u} \big)\Big)}
\end{equation}
where $S_i$ is the set of all the items in the collection, $\lambda$ is the weight of the query in the linear combination. The representations of $Q_0$, $u$ and $i$ will be introduced next. 

\subsection{Query Representation}
In order to generalize the representations to unseen queries, we use the embedding of query words as input and adopt a non-linear projection of the average word embeddings as the representation of a query:
\begin{equation}
\label{eq:fs}
\mathbf{Q_0} = f(\{w_q| w_q \in Q_0\}) = \tanh(W \cdot \frac{\sum_{w_q\in Q_0}\mathbf{w_q}}{|Q_0|} +b)
\end{equation}
where $W \in \mathbb{R}^{d \times d}$ and $b \in \mathbb{R}^{d}$ when the size of embeddings is $d$,  $|Q_0|$ is the length of query $Q_0$.
This method has been shown to be more effective in \cite{ai2017learning} compared with using average embeddings of words and a recurrent neural network to encode the word embedding sequence in the query for product search. 

\subsection{User/Item Language Model} 
To alleviate the potential vocabulary mismatch between queries and items, we also adopt the user/item language model in \cite{ai2017learning} to learn the representation of users and items by constructing language models from their associated reviews. Words in the reviews are assumed to be generated from a multinomial distribution of a user or an item. Take user $u$ for example, given its embedding $\mathbf{u}$ ($\mathbf{u} \in \mathbb{R}^d$) and the embedding of a word w, $\mathbf{w}(\mathbf{w}\in \mathbb{R}^d)$, the probability of $w$ being generated from the language model of $u$ is defined with a softmax function on $\mathbf{w}$ and $\mathbf{u}$: 
\begin{equation}
\label{eq:user_gen}
P(w|u) = \frac{\exp(\mathbf{w}\cdot \mathbf{u})}{\sum_{w' \in S_w}\exp(\mathbf{w'} \cdot \mathbf{u})}
\end{equation}
where $S_w$ is the vocabulary of words in the reviews from the corpus.  
Similarly, the language model for item $i$ is represented with $p(w|i)$, which is the softmax over $\mathbf{w}$ and $\mathbf{i}$. Words are assumed to be generated from the language models of user and items independently.

\subsection{Aspect-Value Generation Model}
\label{subsec:av_gen_model}
We propose an aspect-value generation model, which can be further decomposed to aspect generation given an item and value generation given an aspect and an item.  
Both positive and negative feedback on aspect-value pairs are incorporated into the model.
We first show the assumptions of multivariate Bernoulli distributions for generating aspects and values. Then we show how we construct aspect-value embeddings and learn them in the aspect-value generation model. 

\textbf{Multivariate Bernoulli (MB) Assumption for Aspects.}
We propose a multivariate Bernoulli model for aspect generation. 
Given a purchased item, aspects of the item are assumed to be generated from $n_a$ independent Bernoulli trials of $n_a$ aspects, where $n_a$ is the total number of available aspects and each aspect may have a different probability of appearing in the item's associated aspects. The associated aspects can be any reasonable aspect of the item, e.g., aspects collected from the item's meta-data or reviews. 
Another possible assumption is the multinomial distribution, which is commonly used to model the documents being generated from words in the vocabulary, such as in the query likelihood model \cite{ponte1998language}.
However, this assumption is not appropriate for aspect generation because aspects are not exclusive and the probabilities of all the aspects generated from one item are not necessarily summed to 1. 
For example, for an item, ``style'', ``appearance'', and ``material'' are not mutually exclusive. The higher probability of ``style'' should not lead to the lower probability of ``appearance'' or ``material''.  
So the MB model is more reasonable by considering these aspects generated independently during their own Bernoulli trial. 

\textbf{Multivariate Bernoulli Assumption for Values.}
Similar to aspect generation, the values of an item's aspect are also assumed to be generated from a MB distribution instead of a multinomial distribution. 
The property that probabilities of all the values given an item's aspect are summed to 1 is not suitable, especially for values with negative feedback. 
For example, the aspect, ``battery life'', of an item can be ``short'' or ``terrible'', and a user shows that she does not want the battery life to be short. Minimizing the probability of her ideal item's ``battery life'' to be ``short'' in a multinomial model may lead to a higher probability of ``terrible''. 

Instead of modeling the generation of values with one MB distribution, we propose two independent MB models for the generation of values in positive and negative feedback respectively. 
Positive values are assumed to be generated from $n_v$ independent trials of $n_v$ values and each value has its own probability of appearing in positive values. Negative values are assumed to be generated from a similar process based on its own MB model. 
This approach is more reasonable because values without positive feedback are not necessarily disliked by a user and values on which the user has not provided negative feedback are not necessarily liked.  A value could be valid for the item's aspect but does not receive positive or negative feedback since the system has not asked for feedback on this value, or the user has vague opinions towards the value. 
Our experiments also show better performance of having a separate MB model for negative values compared with using one MB model for both positive and negative values in Section \ref{subsec:ablation}.  

\textbf{Aspect and Value Embeddings}. Words contained in the aspects and values are also in the vocabulary of words in reviews. Since these words represent the characteristics of items, different from words in the reviews that are generated from the item language model, we keep separate embedding lookup tables for the words in the vocabulary of aspects and values to differentiate the properties of same words in the aspect-value pairs or item reviews.  

Aspects of an item can be of multiple words, such as``battery life'' and ``touch screen'',  so we also adopt Equation \ref{eq:fs} and compute the embedding of an aspect $a$ as $\mathbf{a}= f(\{w_a| w_a \in a\})$. 
Positive values and negative values have two separate groups of embeddings, so that values have different representations in the MB models for positive and negative values. 
Since values usually consist of one word, such as ``long'', ``big'', ``clear'', and ``responsive'', the embedding of a value $v$ is just its word embedding, i.e., $\mathbf{v}^+$ for $v$ in the positive values, and $\mathbf{v}^-$ for $v$ in the negative values. Note that these two embeddings are different from the representation of $v$ as a word in the reviews, and values with more than one word were removed from the corpus. 

\textbf{Aspect-Value Probability Estimation.}
Next, we show how to estimate the probabilities in the multivariate Bernoulli models of aspects and values.  
Given the embedding representation of items, aspects and values, the probability of aspect $a$ occurring in the reviews given an item $i$ is 
\begin{equation}
\label{eq:aspect_gen}
P(a \in A(i) |i) =  \delta(\mathbf{a} \cdot \mathbf{i})
\end{equation}
where $A(i)$ is the set of aspects of $i$, and $\delta$ is the sigmoid function $\delta(x) = \frac{1}{1 + e^{-x}}$; 
the probability that value $v$ occurs in the positive value set of item $i$'s aspect $a$, i.e.,  $\{v | I(a,v) = 1\}$, denoted by $V^+(i,a)$, is
\begin{equation}
\label{eq:value_gen}
P(v \in V^{+}(i, a)|i,a) =  \delta(\mathbf{v^+} \cdot (\mathbf{i} + \mathbf{a}))
\end{equation}
where $\mathbf{v^+}$ is the embedding of $v$ as a positive value.
Then the probability that an aspect-value pair $(a,v)$ appears in users' positive feedback given an item $i$ can be computed as:
\begin{equation}
\label{eq:pos_iav}
\begin{aligned}
P(I(a,v)=1| i) &= P(v \in V^{+}(i, a)|i,a)P(a \in A(i) |i) \\
&= \delta(\mathbf{v^+} \cdot (\mathbf{i} + \mathbf{a})) \cdot  \delta(\mathbf{a} \cdot \mathbf{i})
\end{aligned}
\end{equation}
Similarly, the probability that $(a,v)$ occurs in the negative feedback in a conversation that leads to purchasing item $i$, i.e., $P(I(a,v)=-1|i)$ can be calculated according to:
\begin{equation}
\label{eq:neg_iav}
\begin{aligned}
P(I(a,v)=-1| i) &= P(v \in V^{-}(i, a)|i,a)P(a \in A(i) |i) \\
&= \delta(\mathbf{v^-} \cdot (\mathbf{i} + \mathbf{a})) \cdot  \delta(\mathbf{a} \cdot \mathbf{i})
\end{aligned}
\end{equation}
where $V^-(i,a)$ is the set values with negative feedback given $i$ and $a$, and $\mathbf{v^-}$ is the embedding of $v$ as a negative value. 

\subsection{Unified \OurModel~Framework}
With all the components introduced previously, we can learn the embeddings of queries, users, items, aspects and values with a unified framework by maximizing the likelihood of the observed conversations in the training set. For a conversation which was started by user $u$ with an initial request $Q_0$ and leading to a purchased item $i$, under the assumptions of multivariate Bernoulli distributions for aspect and values (Section \ref{subsec:av_gen_model}), we need to consider all the aspects both associated with this conversation and not associated. For each aspect that is associated with the conversation, all the values should be taken into account in the generation of both positive and negative values. Let $ A(i) = \{a|I(a,v)=1\} \cup \{a|I(a,v)=-1\}$ be the aspects that appear in the conversation (same as $A(i)$ in Equation \ref{eq:aspect_gen}), and $S_{a}\setminus A(i)$ be the aspects that have not occurred, where $S_{a}$ is the set of all the aspects in the collection. Let $T_{av}^+=\{(a,v, S_v \setminus \{v\}) | I(a,v) = 1\}$ be the observed instances for positive feedback,  and $T_{av}^-=\{(a,v, S_v \setminus \{v\}) | I(a,v) = -1\}$ be the observed instances for negative feedback, where $S_{v}$ is the set of all the possible values in collection and $S_v \setminus \{v\}$ represents all the values that did not co-occur with the corresponding aspect $a$. The log likelihood of observing the conversation with the reviews of $i$ and $u$, i.e., $R_i$ and $R_u$ respectively, can be computed as 
\begin{equation}
\label{eq:loglikelihood}
\begin{split}
\mathcal{L}(R_i, R_u, u, Q_0, &  S_a \!\setminus \! A(i), T_{av}^+, T_{av}^-, i) \\
&= \log P(R_i, R_u, u, Q_0, S_a \!\setminus \! A(i), T_{av}^+, T_{av}^-, i) \\
\end{split}
\end{equation}
We assume that the probabilities of $R_i$, $R_u$, $S_a \!\setminus \! A(i), T_{av}^+, T_{av}^-$ given $u$, $Q_0$, $i$ are independent. Words in $R_u$ and $R_i$ are supposed to be generated from the language model of $u$ and $i$ respectively. So $R_u$ is independent from $i$ and $Q_0$, and $R_i$ is independent from $u$ and $Q_0$. We also assume that the positive and negative aspect-value instances, $T_{av}^+$ and $T_{av}^-$, only depend on the purchased item $i$. Initial query intent $Q_0$ is considered independent from the user preference $u$. Then Equation \ref{eq:loglikelihood} can be rewritten as: 
\begin{equation}
\label{eq:loglld_detail}
\begin{split}
&\mathcal{L}(R_i, R_u, u, Q_0, S_a \!\setminus \! A(i), T_{av}^+, T_{av}^-, i) \\
&= \log P(R_i, R_u, S_a \!\setminus \! A(i), T_{av}^+, T_{av}^-| u, Q_0, i) P(u,Q_0,i)\\
&= \log \big( P(R_u|u) P(R_i|i)  \\
& \;\;\;\;\;\;\;\;\;\;\; P(S_a \!\setminus \! A(i)|i) P(T_{av}^+|i) P(T_{av}^-|i)  P(i|u,Q_0) P(u)P(Q_0)\big) \\
&\simeq \log P(i|u,Q_0) + \sum_{w \in R_i}\!\! \log P(w|i) + \!\!\! \sum_{w \in R_u}\!\! \log P(w|u) \\
& \;\;\;\;\; + \!\!\!\!\!\!\sum_{a\in S_a \!\setminus \! A(i)} \!\!\!\!\!\!\log \big(1- P(a \in A(i)|i)\big) + \log P(T_{av}^+|i) + \log P(T_{av}^-|i) \hspace{-30pt}\\
\end{split}
\end{equation}
$P(u)$ and $P(Q_0)$ are predefined as uniform distributions, and thus ignored in the equation.  
$P(T_{av}^+|i)$ and $P(T_{av}^-|i)$ can be computed in a similar way. Take $\log P(T_{av}^+|i)$ for instance, we can compute it as:
\begin{equation}
\label{eq:av_likelihood}
\begin{split}
& \log P(T_{av}^+|i)= \!\!\!\!\!\!\sum_{(a,v,\mathcal{V}) \in T_{av}^+} \!\!\!\!\!\! \big( \log P(v, \mathcal{V}|a,i) + \log P(a\in A(i)|i) \big)\\
& \;\;\;\;\;  = \!\!\!\!\!\! \sum_{(a,v,\mathcal{V}) \in T_{av}^+} \!\!\!\! \Big(  \log P(a \in A(i)|i) + \log P(v\in V^+(a,i)|a,i)  \\
& \;\;\;\;\;\;\;\;\;\;\;\;\;\;\;\;\;\;\;\;\;\;\;\;\;\;\;  + \sum_{v'\in \mathcal{V}}\big(1- P(v' \in V^+(a,i)|a,i)\big) \Big) \\
\end{split}
\end{equation}
where $\mathcal{V}\! =\! S_v \!\setminus\! \{v\}$ and $V^+(a,i)$ is the set of positive values associated with $a$ and $i$. $P(T_{av}^-|i)$ can be computed with $V^+(a,i)$ replaced by $V^-(a,i)$, i.e., the set of negative values corresponding to aspect $a$ of $i$.
From Equation \ref{eq:loglld_detail} \& \ref{eq:av_likelihood}, the overall log likelihood of an observed conversation is the sum of the log likelihood for the user language model, item language model, item generation model, aspect generation model and value generation model.

It is impractical to compute the log likelihood directly since it involves softmax function to compute the probability (Equation \ref{eq:user_gen} and \ref{eq:item_gen}), which has the sum of a large number of elements as the denominator. Same as \cite{ai2017learning}, we adopt the negative sampling strategy to approximate the estimation of the softmax function. Specifically, $\beta$ random samples are randomly selected from the corpus according to a predefined distribution and used as negative samples to approximate the denominator of the softmax function. So the log likelihood of the user language model with negative sampling is:
\begin{equation}
\label{eq:neg_ulang}
\log P(w|u) = \log \delta (\mathbf{u}\cdot \mathbf{w}) 
+ \beta \cdot \mathbb{E}_{w' \sim P_w}[ \log \delta (-\mathbf{u}\cdot \mathbf{w'})]
\end{equation}
where $P_w$ is defined as the word distribution in the reviews of the corpus, raised to $\frac{3}{4}$ power \cite{mikolov2013distributed}. The log likelihood of the item language model can be approximated with $u$ replaced by $i$ in Equation \ref{eq:neg_ulang}. Similarly, the log likelihood of the item generation model is computed as:
\begin{equation}
\begin{aligned}
\log &P(i|u,Q_0) = \log \delta \Big( \mathbf{i} \cdot \big(\lambda \mathbf{Q_0} + (1-\lambda) \mathbf{u}\big )\Big) \\
& + \beta \cdot \mathbb{E}_{i' \sim P_i} \Big [\log \delta\Big( -\mathbf{i'} \cdot (\lambda \mathbf{Q_0} + (1-\lambda) \mathbf{u})\Big) \Big]
\end{aligned}
\end{equation}
where $P_i$ is predefined as a uniform distribution for items. 

Since the sets of aspects and values, namely $S_{a}$ and $S_{v}$, are usually large but the number of aspects and values that appear in a conversation is small, it would be inefficient to consider the whole set of $S_a \! \setminus \!A(i)$ and $\mathcal{V}$ (i.e., $S_v \!\setminus \! \{v\}$) in Equation \ref{eq:loglld_detail} and \ref{eq:av_likelihood}. We random selected $\beta$ samples from $S_{a} \! \setminus \! A(i)$ and $\mathcal{V}$ to represent the whole set. 

The final objective of our model is to optimize the log likelihood of all the conversations in the training set together with L2 regularization to avoid overfitting, i.e., 
\begin{equation}
\begin{split}
\label{eq:final_loss}
&\mathcal{L'} = \sum_{u, Q_0, i} \mathcal{L}(R_i, R_u, u, Q_0, S_a \!\setminus \! A, T_{av}^+, T_{av}^-, i) \\
& + \gamma \big(\!\! \sum_{w \in S_w}\!\!\! \mathbf{w}^2 \!\! +\!\!  \sum_{u \in S_u} \!\!\! \mathbf{u}^2 \!\! + \!\! \sum_{i \in S_i}
\!\!\! \mathbf{i}^2 + \sum_{a \in S_a} \!\!\!\mathbf{a}^2 \!\! + \!\!  \sum_{v \in S_v} \!\!\! (\mathbf{v^+})^2 \!\!  + \!\! \sum_{v \in S_v} \!\!\! (\mathbf{v^-})^2 \big)
\end{split}
\end{equation}
where $S_u$ is the set of users, $\gamma$ is the coefficient for L2 regularization, $\mathbf{v^+}$ and $\mathbf{v^-}$ are the embeddings of $v$ as a positive value and as a negative value respectively, kept in two different lookup tables. 
All the embeddings are trained simultaneously in our model. 

\subsection{Item Ranking with \OurModel}
After we get the embeddings of words, users, items, aspects and values as positive or negative targets, when a user $u$ issues a new query $Q_0$, in the first iteration, our system ranks an item $i$ based on $P(i|u, Q_0)$ according to Equation \ref{eq:item_gen}. In the $k$-th iteration ($k>1$) of the conversation, besides $u$ and $Q_0$, the positive and negative feedback on aspect-value pairs collected in previous $k-1$ iterations also act as the basis for ranking. Let $AV^+$ and $AV^-$ be the aspect-value pairs with positive and negative feedback respectively, item $i$ is ranked according to
\begin{equation}
\label{eq:rank_func}
\begin{split}
\log &P(u,Q_0, AV^+, AV^-|i) = \log \frac{P(AV^+, AV^-|u,Q_0,i)P(u,Q_0,i)}{P(i)} \hspace{-30pt} \\
&= \log \frac{P(AV^+|i)P(AV^-|i) P(i|u,Q_0) P(u)P(Q_0)} {P(i)} \\
&\stackrel{rank}{=} \!\!\!\! \sum_{(a,v) \in AV^+} \!\!\!\!\!\!\!\!\log \Big(\delta \big(\mathbf{v^+} \cdot (\mathbf{i} + \mathbf{a}) \big) \cdot  \delta(\mathbf{a} \cdot \mathbf{i})\Big) \\
& \;\;\;\;\;\;\;+ \!\!\!\!\!\!\!\!\!\!  \sum_{(a,v)\in AV^-}\!\!\!\!\!\!\!\!\log \Big( \delta\big(\mathbf{v^-} \cdot (\mathbf{i} + \mathbf{a})\big) \cdot  \delta(\mathbf{a} \cdot \mathbf{i}) \Big)
+   \mathbf{i} \cdot \big(\lambda \mathbf{Q_0} + (1-\lambda) \mathbf{u} \big) \hspace{-30pt} \\
\end{split}
\end{equation}
The inference process is simple, so we omit it due to space limit. The time complexity for item ranking is $O(md|S_i|)$, where $m$ is the number of aspect-value pairs used for re-ranking, $d$ is the embedding size, and $|S_i|$ is total number of items in the corpus. 

\section{Experimental Setup}
\label{sec:exp_setup}
In this section, we introduce our experimental settings. 
We first introduce the dataset and evaluation methodology for our experiments. Then we describe the baseline methods and training settings for our model.
\subsection{Datasets}
\begin{table}
	\caption{Statistics of Amazon datasets.}
	\centering
	\label{tab:stats}  
	\small
	\begin{tabular}{l  r  r  r }
		\hline
		\multirow{2}{*}{Dataset} & Health \&  & Cell Phones \&  & Movies \&  \\
		&  Personal Care &  Accessories &  TV \\
		\hline
		\#Users & 38,609 & 27,879 & 123,960 \\
		\#Items & 18,534 & 10,429 & 50,052 \\
		\#Reivews & 346,355 & 194,439 & 1,697,524 \\
		\#Queries & 779 & 165 & 248 \\
		Query length & 8.25$\pm$2.16 & 5.93$\pm$1.57 & 5.31$\pm$1.61 \\
		\hline
		\#Aspects & 1,906 & 738 & 6,694 \\
		\#Values & 1,988 & 1,052 & 6,297 \\
		\#AV pairs & 15,297 & 7,111 & 82,060 \\
		\hline
		\multicolumn{4}{l}{\#User-query pairs} \\
		Train & 231,186 & 114,177 & 241,436 \\
		Test  & 282 & 665 & 5,209 \\
		\hline
		\multicolumn{4}{l}{\#Rel items per user-query pair} \\
		Train & 1.14$\pm$0.48 & 1.52$\pm$1.13 & 5.40$\pm$18.39 \\
		Test & 1.00$\pm$0.00 & 1.00$\pm$0.05 & 1.10$\pm$0.49 \\
		\hline
	\end{tabular}
\end{table}

\label{subsec:datasets}
\textbf{Dataset Description.} As in previous research on product search \cite{ai2017learning,van2016learning, zhang2018towards}, we also adopt the Amazon product dataset \cite{mcauley2015inferring} for experiments. There are millions of customers and products as well as rich meta-data such as reviews, multi-level product categories and product descriptions in the dataset. We used three categories in our experiments, which are \textit{Movies \& TV}, \textit{Cell Phones \& Accessories} and \textit{Health \& Personal Care}. The first one is large-scale while the rest two are smaller. We experimented on these datasets to see whether our model is effective on collections of different scales. The statistics of our datasets are shown in Table \ref{tab:stats}. Since there are no datasets that have the sequence of 
$u\rightarrow Q_0;D_1,\mathcal{Q}_{1,1},I_{1,1}, \cdots,$ $\mathcal{Q}_{1,m}, I_{1,m} \cdots,D_k,\mathcal{Q}_{k,1},I_{k,1}, \cdots, \mathcal{Q}_{k,m}, I_{k,m} \rightarrow i$ as a conversation during product search, we need to construct such conversations for the datasets.

\textbf{Initial Query Construction.}
To construct initial queries $Q_0$ in the conversation, we adopt the three-step paradigm of extracting queries for each item, same as the previous work \cite{van2016learning, ai2017learning, zhang2018towards}. First, the multi-level category information of each item is extracted from the meta-data. Then, the terms in the categories are concatenated to form a topic string. At last, stopwords and duplicate words are removed. In this way, there can be multiple queries extracted for each item. 
When a user purchased an item, all the queries associated with the item can be considered as the initial query which is issued by the user that finally leads to purchasing the item. The queries extracted are general and do not reveal specific information of the purchased items.
Examples queries are ``health personal care dietary supplement vitamin'', ``cell phone accessory international charger'', ``tv movies'' for each category.  

\textbf{Conversation Construction.}
The essential part to construct a conversation for a user-query pair is to extract the aspect-value pairs from the items. We adopt the aspect-value pair extraction toolkit by \citet{zhang2014explicit,zhang2014users} to extract the pairs from the reviews of the items in each dataset. 
During training, random items were selected as non-relevant results for a user-query pair ($u, Q_0$) since few items are relevant among the entire collection. Then all the aspect-value pairs extracted from the non-relevant items were used to form corresponding questions. During test time, the aspect-value pairs that were mentioned most in the non-relevant items retrieved in the previous iterations were selected to formulate questions.
Table \ref{tab:av_example} shows some common aspect-value pairs extracted from the reviews of an item which corresponds to the example query. In contrast to facets based on which filtering can be applied \cite{lim2010multi,vandic2013facet}, our extracted aspects and values are more flexible and not exclusive, which makes simple filtering not reasonable. 
During the conversation, positive or negative feedback on the aspect-value pairs can be constructed. 

Previous works \cite{zhang2018towards, Sun:2018:CRS:3209978.3210002} on conversational search and recommendation construct users' response to the system's questions according to their ideal items, which show their hidden intent. 
In their experiments, the system asks users their preferred values of an aspect and answers are constructed according to their purchased items or their reviewed restaurants. We also simulate user feedback following the same paradigm. 
For a question on an aspect-value pair, when the aspect matches an aspect extracted from the purchased item $i$, if their values also match, the aspect-value pair is considered to have positive feedback, otherwise, the pair is assumed to receive negative feedback. If the aspect in the question does not match any aspect associated with $i$, no answers are collected from users.


\begin{table}
	\caption{Examples of extracted aspect-value pairs. }
	\centering
	\label{tab:av_example}
	\begin{tabular}{l l l}
		\hline
		Query & Aspect & Value \\
		\hline
		& color & white, black, pink, red \\	
		& fit & snug, loose \\
		cell phone accessory & material & plastic, rubbery \\	
		waterproof case & plastic & soft, hard, thin, thick \\	
		& case & flimsy, protective, sturdy \\		
		& cover & dark, clear \\
		\hline
	\end{tabular}

\end{table}

\subsection{Evaluation Methodology}
\label{subsec:eval_method}
As in \cite{ai2017learning}, we randomly select 70\% of the reviews for each user in the training set and keep the other 30\% in the test set. Each review indicates that a user purchases a corresponding item. Then 30\% of all the available queries are divided into the test set. If for an item in a training set, all it associated queries are in the test set, we randomly move one query back to the training set. This assures that each item has at least one query in the training data and each tuple of user, query, purchased item in the test set is not observed in the training set. Finally, all the available user-query pairs in the test set are used to test the performance of the corresponding conversations.
Statistics of train/test splits can be found in Table \ref{tab:stats}.

To evaluate the performance of the models in the first $k$-th iterations in a conversation, we use the freezing ranking paradigm \cite{ruthven2003survey,bi2018iterative}, which is commonly used for evaluating relevance feedback, to maintain a rank list. Items shown to the user in the previous $k-1$ iterations are frozen, and the remaining items are re-ranked and appended to the frozen items to form the rank list of all the items.
Note that our system does not need to show a long list to the user in each iteration; 
we keep the items which are not shown in the conversations in the rank lists to avoid that most methods have nearly zero scores for the evaluation metrics. Besides, whenever a relevant item is retrieved in the previous iterations, the ranking of all the items will not be updated in the following iterations. For models that do not utilize feedback, the evaluation is based on the rank lists retrieved with $u$ and $Q_0$.

Mean average precision ($MAP$) and mean reciprocal rank ($MRR$) at cutoff 100, as well as normalized discounted cumulative gain ($NDCG$) at 10 are used to evaluate the rank lists in each iteration. $MRR$ indicates the average iterations the system needs to find a relevant item. $MAP$ measures the overall performance of a system in terms of both precision and recall. $NDCG@10$ focuses on the performance of the system to retrieve relevant results in the first 10 iterations, especially in earlier iterations. 

\subsection{Baselines}
\label{subsec:baseline}
We compare our aspect-value based embedding model with three groups of baselines, which are word and embedding based retrieval models that do not consider feedback, and models using item-level negative feedback.

\textbf{BM25.} 
BM25 \cite{robertson1995okapi} scores a document according to a function of the term frequency, inverse document frequency of query terms and document length. 

\textbf{QL.}
The query likelihood model (QL) \cite{ponte1998language} ranks a result according to the log-likelihood that the query words are generated from the unigram language model of the result.

\textbf{LSE.}
The latent semantic entity (LSE) model \cite{van2016learning} is a non-personalized product search model, which learns the vectors of words and items by predicting the items with n-grams in their reviews. 

\textbf{HEM.}
The hierarchical embedding model (HEM) \cite{ai2017learning} is a state-of-art personalized product search model that \OurModel~ is based on. It has the item generation model and language models of users and items. We use the best version reported in \cite{ai2017learning} which uses non-linear projected mean for query embeddings and set the query weight $\lambda = 0.5$ (in Equation \ref{eq:item_gen}) in both HEM and our own model. 

\textbf{Rocchio.}
Only the part of moving query model further from non-relevant results in Rocchio \cite{rocchio1971relevance} takes effect in our scenario since only non-relevant results are available. BM25 \cite{robertson1995okapi} function is used for weighting terms. 

\textbf{SingleNeg.}
SingleNeg \cite{karimzadehgan2011improving} extracts a single negative topic model from a batch of non-relevant results by considering they are generated from the mixture of the language model of the negative topic and the background corpus. The negative topic model is then used to adjust the initial relevance score. 

\textbf{MultiNeg.}
MultiNeg \cite{karimzadehgan2011improving} considers that each non-relevant result is generated from a corresponding negative topic model and use multiple negative models to adjust the original relevance score. 

BM25 and QL are word-based retrieval models. LSE and HEM are embedding-based models for non-personalized and personalized product search. Rocchio, SingleNeg, and MultiNeg incorporate item-level negative feedback collected from previous iterations. For the initial ranking, we use BM25 for Rocchio, QL for SingleNeg and MultiNeg respectively. We get the performance of BM25 and QL using galago\footnote{\url{https://www.lemurproject.org/galago.php}} with default parameter settings. We implemented Rocchio, SingeNeg and MultiNeg based on galago and tuned the term count for negative model from $\{10,20,30,40,50\}$, the weight for negative documents from $\{0.01, 0.05, 0.1, 0.2,0.3,0.4\}$.

\subsection{Model Parameter Settings}
\label{subsec:model_train}
We implemented our model and HEM with PyTorch 
\footnote{https://pytorch.org/} 
and LSE with Tensorflow
\footnote{https://www.tensorflow.org/}.
LSE, HEM and our model are all trained with stochastic gradient descent for 20 epochs with batch size 64. Initial learning rate is set to 0.5 and gradually decrease to 0 during training. The gradients with global norm larger than 5 were clipped to avoid unstable updates. To reduce the effect of common words, as in \cite{mikolov2013distributed, ai2017learning}, we set the sub-sampling rate of words as $10^{-5}$ for \textit{Cell Phones \& Accessories} and \textit{Health \& Personal Care}, and $10^{-6}$ for \textit{Movies \& TV}. L2 regularization strength $\gamma$ was tuned from 0.0 to 0.005. 
The embedding size $d$ was scanned from $\{100,200, \cdots, 500\}$. The effective of embedding size will be shown in Section \ref{subsec:para_sensitivity}. Negative samples $\beta$ in Equation \ref{eq:neg_ulang} \& \ref{eq:neg_iav} were set to 5. For conversation construction during training, 2 random items were sampled as non-relevant results and all the positive and negative values with matched aspects were used in the conversation. 
For testing, the total number of iterations for retrieval in the conversation was set from 1 to 5. In the first iteration, there is no feedback collected. During each iteration, the number of aspect-value pairs, on which the feedback is provided, namely, $m$ in Section \ref{eq:conv_seq}, is selected from $\{1,2,3\}$. 
We only report the results of the best settings for all the methods in Section \ref{sec:results}.


\section{Results and Discussion}
\label{sec:results}
In this section, we discuss the results of our experiments. 
We first compare the overall retrieval performance of both \OurModel~ and the state-of-the-art product search baselines in Section \ref{subsec:overall_perf}. 
Then we study the effect of different model components, feedback processes, and embedding sizes on each model in the following subsections.

\begin{table*}
	\caption{Comparison between baselines and our model \OurModel.  Numbers marked with`$^*$' are the best baseline performances. `$^+$' indicates significant differences between the iterative feedback models and their corresponding initial rankers in Fisher random test \cite{smucker2007comparison} with $p<0.05$, i.e., Rocchio vs BM25, SingleNeg and MultiNeg vs QL, \OurModel$_{pos}$, \OurModel$_{neg}$ and \OurModel$_{all}$ vs \OurModel$_{init}$. `$\dagger$' denotes significant improvements upon the best baseline. Best performances are in bold. } 
	\label{tab:overallperf}
	\large
	\scalebox{0.92}{    
		\begin{tabular}{  l || l | l | l || l | l | l || l | l | l   }
			\hline
			Dataset& \multicolumn{3}{c||}{Health \& Personal Care} & \multicolumn{3}{c||}{Cell Phones \& Accessories} & \multicolumn{3}{c}{Movies \& TV} \\
			\hline
			Model & $MAP$ & $MRR$ & $NDCG$ & $MAP$ & $MRR$ & $NDCG$ & $MAP$ & $MRR$ & $NDCG$ \\
			\hline
			\hline
			BM25 & 0.055 & 0.055 & 0.053 & 0.065 & 0.065 & 0.077 & 0.012 & 0.009 & 0.008 \\
			\hline
			Rocchio & 0.055 & 0.055 & 0.053 & 0.065 & 0.065 & 0.077 & 0.012 & 0.009 & $0.009^+$ \\
			\hline
			\hline
			QL & 0.046 & 0.046 & 0.048 & 0.063 & 0.062 & 0.076 & 0.016 & 0.012 & 0.015 \\
			\hline
			SingleNeg & 0.046 & 0.046 & 0.048 & 0.063 & 0.062 & 0.076 & $0.018^+$ & $0.015^+$ & $0.017^+$ \\
			MultiNeg & 0.046 & 0.046 & 0.048 & 0.063 & 0.062 & 0.076 & $0.018^+$ & $0.015^+$ & $0.016^+$ \\
			\hline
			\hline
			LSE & 0.155 & 0.157 & 0.195 & 0.098 & 0.098 & 0.084 & 0.023 & 0.025 & 0.027 \\
			HEM & $0.189^*$ & $0.189^*$ & $0.201^*$ & $0.115^*$ & $0.115^*$ & $0.116^*$ & $0.026^*$ & $0.030^*$ & $0.030^*$ \\
			\hline
			\hline
			\OurModel$_{init}$ & $0.227^{\dagger}$ & $0.227^{\dagger}$ & $0.233^{\dagger}$ & $0.126^{\dagger}$ & $0.126^{\dagger}$ & $0.130^{\dagger}$ & $0.028^{\dagger}$ & $0.030$ & $0.031^{\dagger}$ \\
			\hline
			\OurModel$_{pos}$ & 0.225$^\dagger$ & 0.225$^\dagger$ & 0.250$^{+\dagger}$ & 0.133$^{+\dagger}$ & 0.133$^{+\dagger}$ & 0.135$^{+\dagger}$ & 0.031$^{+\dagger}$ & 0.033$^{+\dagger}$ & 0.035$^{+\dagger}$ \\
			\OurModel$_{neg}$ & \textbf{0.260$^{+\dagger}$} &  \textbf{0.260$^{+\dagger}$} &  \textbf{0.305$^{+\dagger}$} &  \textbf{0.154$^{+\dagger}$} &  \textbf{0.154$^{+\dagger}$} &  \textbf{0.177$^{+\dagger}$} &  0.033$^{+\dagger}$ &  0.035$^{+\dagger}$ &  0.038$^{+\dagger}$ \\
			\OurModel$_{all}$ & 0.236$^{+\dagger}$ & 0.236$^{+\dagger}$ & 0.258$^{+\dagger}$ & 0.145$^{+\dagger}$ & 0.145$^{+\dagger}$ & 0.145$^{+\dagger}$ & \textbf{0.034$^{+\dagger}$} &  \textbf{0.036$^{+\dagger}$} &  \textbf{0.042$^{+\dagger}$} \\
			\hline
			\hline 
			
		\end{tabular}
	}
\end{table*}

\subsection{Overall Retrieval Performance}
\label{subsec:overall_perf}
Table \ref{tab:overallperf} shows the retrieval performance of all the methods in the conversational product search on different Amazon sub-datasets (i.e., \textit{Movies \& TV}, \textit{Cell Phones \& Accessories} and \textit{Health \& Personal Care}).
Specifically, we use BM25 and QL as the initial models to generate the first-round retrieval results for Rocchio and SingNeg/MultiNeg, respectively.
Also, we refer to the \OurModel~without feedback, with positive feedback, with negative feedback, and with both positive and negative feedback on aspect-value pairs as \OurModel$_{init}$, \OurModel$_{pos}$, \OurModel$_{neg}$, and \OurModel$_{all}$, respectively. 

As shown in Table~\ref{tab:overallperf}, term-based retrieval models perform worse than neural embedding models.
Without feedback information, QL and BM25 are approximately 50\% worse than LSE and HEM on all datasets in our experiments.
As discussed by previous studies~\cite{van2016learning,ai2017learning}, there are no significant correlations between user purchases and the keyword matching between queries and product reviews.
Thus, term-based retrieval models usually produce inferior results in product search.
Among different embedding-based product retrieval models, \OurModel$_{init}$ achieves the best performance and significantly outperforms HEM and LSE on all the three datasets. 
This indicates that incorporating aspect-value information into search optimizations is generally beneficial for the performance of product search systems. 
\footnote{MRR and MAP are almost the same for Health \& Personal Care and Cell Phones \& Accessories since users purchase only 1 item under each query most of time in these categories (see Table \ref{tab:stats}). }

After a 5-round iterative feedback process, we observe different results for different feedback models.
For term-based negative feedback models such as Rocchios, SingleNeg, and MultiNeg, we observe little performance improvement during the feedback process.
Comparing to their initial retrieval models in the first iteration (i.e., BM25 and QL), term-based feedback models only achieve significant MRR improvements on \textit{Movies \& TV}.
For \OurModel, on the other hand, we observe consistent and large improvements over the initial retrieval model (i.e., \OurModel$_{init}$) in all three datasets.
The performance of the best \OurModel~is approximately 10\% to 20\% better than \OurModel$_{init}$ in terms of MRR.

Among different variations of \OurModel, \OurModel$_{neg}$ performs the best on \textit{Cell Phones \& Accessories} and \textit{Health \& Personal Care}, while \OurModel$_{all}$ performs the best on \textit{Movies \& TV}. 
Overall, it seems that negative aspect-value feedback tends to provide more benefits for \OurModel~than positive aspect-value feedback. 
In a positive feedback scenario, feedback information is ``inclusive''.
In other words, all aspect-value pairs from relevant items could be used to generate positive feedback, but this does not mean that all relevant items should have the same property.
For example, a user who tells the system to find a ``red'' phone case may also be satisfied with a ``pink'' phone case.
In contrast, in a negative feedback scenario, feedback information is ``exclusive''.
When a user says ``I don't like red'', it means that any items with color ``red'' is definitely not relevant to this user.
Thus, negative feedback information could be more useful for the filtering of irrelevant products. 

\subsection{Ablation Study} 
\label{subsec:ablation}

\begin{figure}
	\centering
	\includegraphics[width=2.3in]{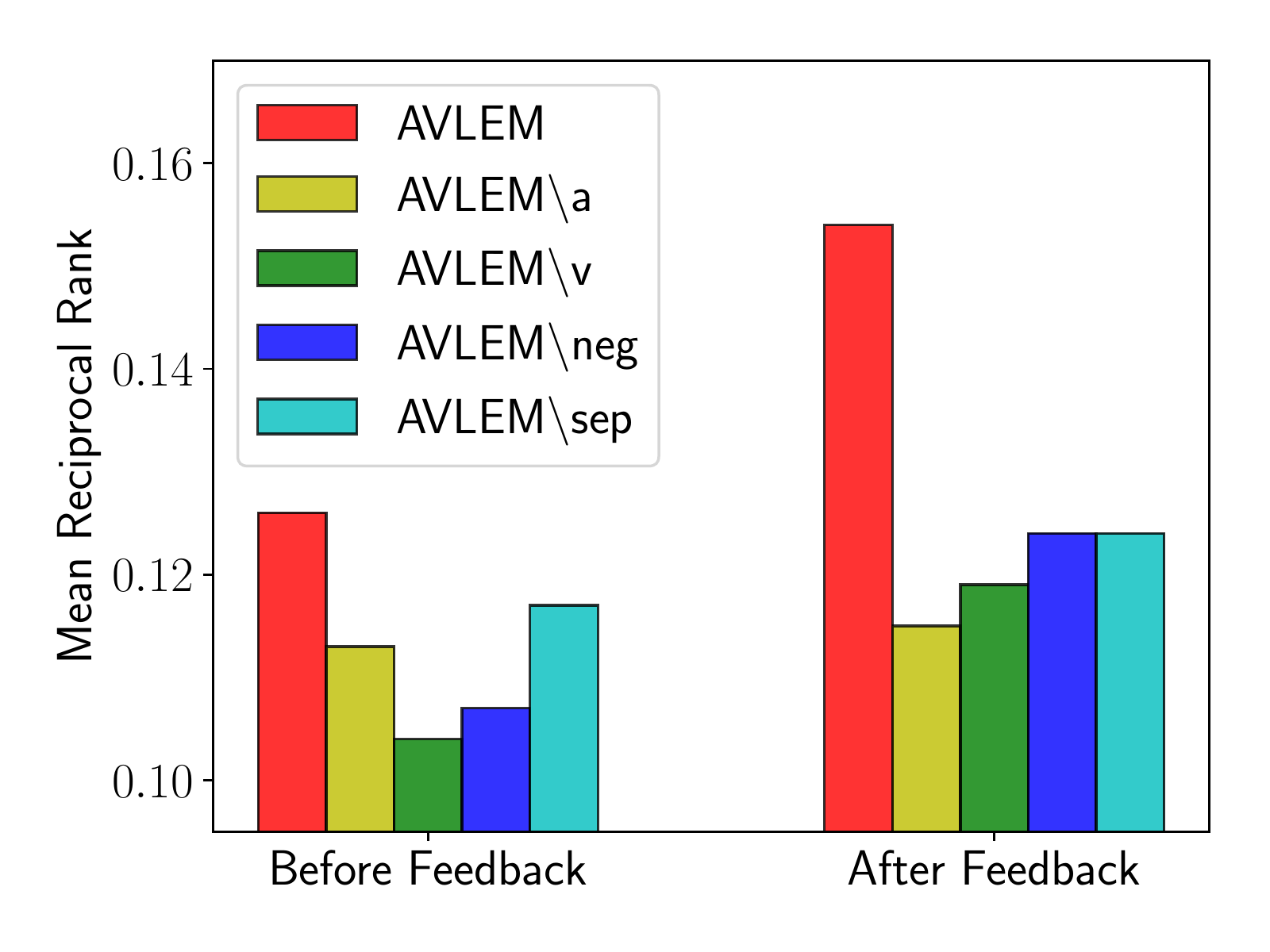}
	\caption{The MRR of \OurModel~ with different components removed on \textit{Cell Phones \& Accessories}.}
	\label{fig:ablation_mrr}
\end{figure}

\begin{figure*}
	\centering
	\begin{subfigure}{.33\textwidth}
		\includegraphics[width=2.22in]{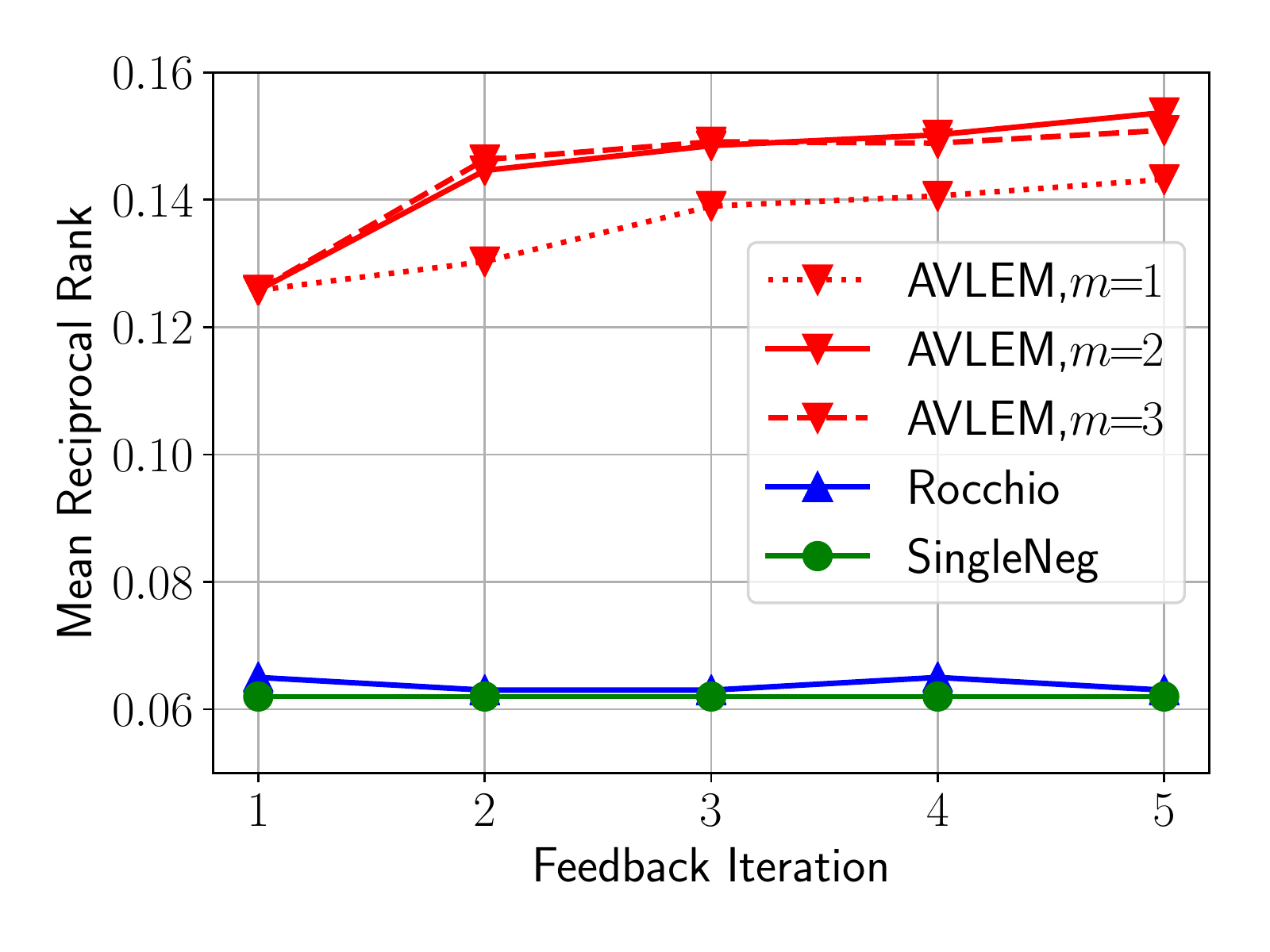}
		\caption{Feedback Performance}
		\label{fig:av_num_mrr}
	\end{subfigure}%
	\begin{subfigure}{.33\textwidth}
		\includegraphics[width=2.22in]{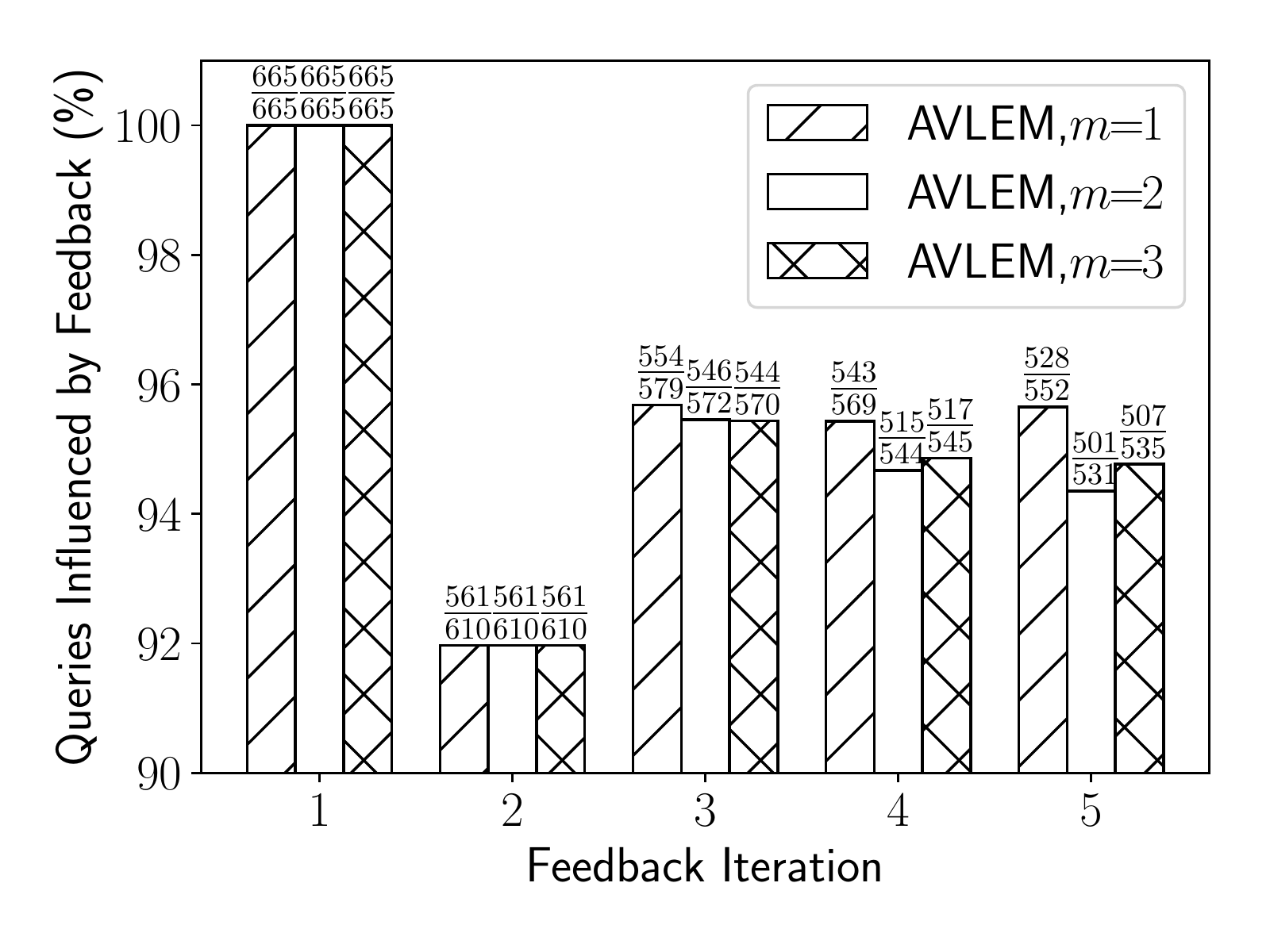}
		\caption{Feedback Query Coverage}
		\label{fig:av_num_query_coverage}
	\end{subfigure}%
	\begin{subfigure}{.33\textwidth}
		\includegraphics[width=2.22in]{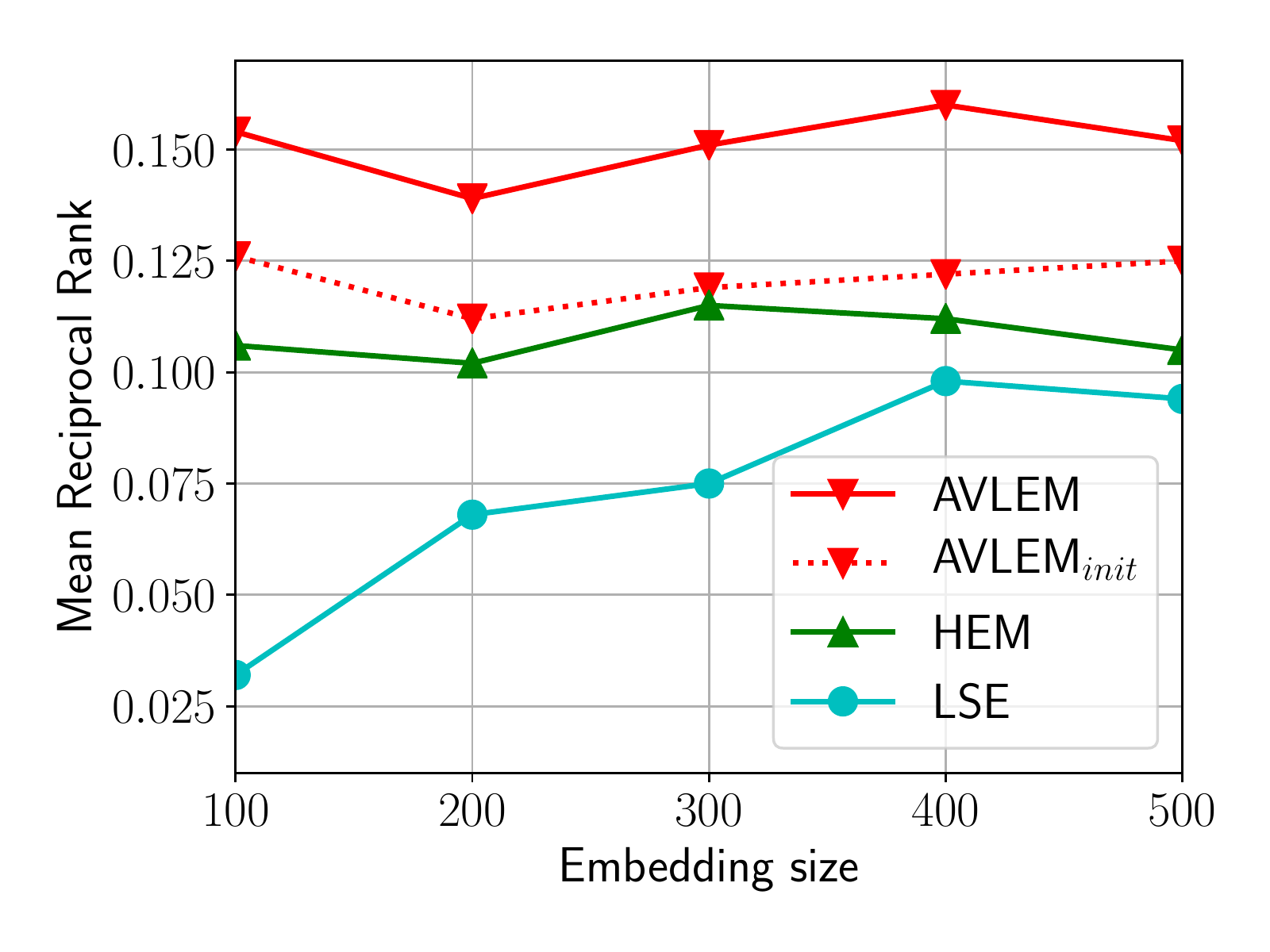}
		\caption{Effect of Embedding Sizes}
		\label{fig:embed_mrr}
	\end{subfigure}%
	\caption{The parameter sensitivity analysis of baselines and \OurModel~on \textit{Cell Phones\&Accessories}.}

\end{figure*}
In order to evaluate the importance of different model components, we conduct ablation experiments by removing the aspect generation network (i.e., $P(a\in A(i) | i)$ in Equation~ \ref{eq:loglld_detail} \& \ref{eq:av_likelihood}), the value geneartion network (i.e., $P(v \in V^{+/-}(a,i)|a, i)$ in Equation~\ref{eq:av_likelihood}), or the negative feedback network (i.e., $P(T_{av}^-|i)$ in Equation \ref{eq:loglld_detail}) for \OurModel.
We refer them as \OurModel$ \setminus $a, \OurModel$ \setminus $v, and \OurModel$\setminus$neg, respectively. 
Also, we refer to the \OurModel~ that uses a single set of value embedding representations for both $\mathbf{v}^+$ and $\mathbf{v}^-$ in Equation~\ref{eq:final_loss} as \OurModel$\setminus$sep.
In \OurModel$\setminus$neg and \OurModel$\setminus$sep, we do not have a separate embedding representations for $v \in V^{-}(a,i)$ in $P(v \in V^{-}(a,i)|a, i)$.
Instead, we replace $P(v \in V^{-}(a,i)|a, i)$ with $1 - P(v \in V^{+}(a,i)|a, i)$ in Equation~\ref{eq:final_loss} to train and test these two models.

Figure~\ref{fig:ablation_mrr} depicts the performance of \OurModel~with different components removed on \textit{Cell Phones \& Accessories}.
We group the results here into two categories -- the model performance before feedback (i.e., \OurModel$_{init}$) and the model performance after feedback (i.e., \OurModel).
As shown in the figure, removing $P(v \in V^{+/-}(a,i)|a, i)$ in Equation~ \ref{eq:av_likelihood} (i.e., \OurModel$ \setminus $v) results in a significant drop of retrieval performance for \OurModel~ before feedback, which means that the relationships between items and aspect-values are important for effectively learning item representations in product search.
Also, without the aspect generation model $P(a\in A(i) | i)$, we observe almost no performance improvement on \OurModel$\setminus$a after the incorporation of feedback information. 
This indicates that understanding the relationships between items and product aspects are crucial for the use of aspect-value based feedback signals.
Last but not least, we notice that both the removing of $P(T_{av}^-|i)$ in Equation \ref{eq:loglld_detail} (i.e., \OurModel$ \setminus $neg) and the unifying of item embeddings in positive and negative feedback (i.e., \OurModel$ \setminus $sep) lead to inferior retrieval performance before and after feedback.
As discussed in Section~\ref{subsec:av_gen_model}, the use of negative aspect-value pairs and the separate modeling of value embedding in different feedback scenarios are important for the multivariate Bernoulli assumptions.
By replacing $P(v \in V^{-}(a,i)|a, i)$ with $1 - P(v \in V^{+}(a,i)|a, i)$, we jeopardize the foundation of \OurModel, which consequentially damages its retrieval performance in our experiments.

\subsection{Parameter Sensitivity}
\label{subsec:para_sensitivity}


\textbf{Effect of the Amount of Feedback.}
There are two important hyper-parameters that control the simulation of conversational feedback in our experiments: the number of feedback iterations and the number of product aspects in each iteration ($m$).
Figure~\ref{fig:av_num_mrr} depicts the performance of different feedback models with respect to feedback iterations on \textit{Cell Phones \& Accessories}. 
As shown in the figure, the performance of Rocchio and SingleNeg does not show any significant correlations with the increasing of feedback iterations. 
In contrast, the performance of \OurModel~ gradually increases when we provide more feedback information. 
The MRR of \OurModel~ with 1 product aspect per iteration improves from 0.126 to 0.143 after 5 rounds of feedback.
Also, \OurModel~ generally achieves better performance when we increase the number of feedback aspects from 1 to 3.
This indicates that our model can effectively incorporate feedback information in long-term conversations.

To further analyze the effect of multi-iteration feedback, we show the percentage of queries influenced by \OurModel~ in each iteration on \textit{Cell Phones \& Accessories} in Figure~\ref{fig:av_num_query_coverage}.
Notice that iteration 1 represents the initial retrieval of the feedback process, and this is the reason when all queries are affected by \OurModel.
As we can see, the percentages of influenced queries remain roughly unchanged (from 92\% to 96\%) after each feedback iteration. 
This means that feedback aspects have been effectively generated by our simulation process in most cases. 
Also, during the feedback process, the number of available test queries (i.e., the queries with no relevant items retrieved in the previous iterations) gradually decreases from 665 to 531 for the best \OurModel~ (i.e., \OurModel~ with 2 product aspects per feedback iteration), which means that more and more relevant items have been retrieved.



\textbf{Effect of Embedding Size.}
Figure \ref{fig:embed_mrr} shows the sensitivity of both our models and the neural product retrieval baselines (i.e., HEM and LSE) in terms of embedding size on \textit{Cell Phones \& Accessories}. 
While we observe a slight MRR improvement for LSE after increasing the embedding sizes from 100 to 500, we do not see similar patterns for both HEM and \OurModel. 
Also, the performance gains obtained from the feedback process for our model (\OurModel~v.s.\OurModel$_{init}$) are stable with respect to the changes of embedding sizes.


\section{Conclusion and Future Work}
\label{sec:conclusion}
In this paper, we propose a paradigm for conversational product search based on negative feedback, where the system identifies users' preferences by showing results and collecting feedback on the aspect-value pairs of the non-relevant items. To incorporate the fine-grained feedback, we propose an aspect-value likelihood model that can cope with both positive and negative feedback on the aspect-value pairs. 
It consists of the aspect generation model given items and value generation model given items and aspects. 
One multivariate Bernoulli (MB) distribution is assumed for the aspect generation model, and two other MB distributions are assumed for the generation of positive and negative values. 
Experimental results show that our model significantly outperforms the state-of-the-art product search baselines without using feedback and baselines using item-level negative feedback. 
Our work has the limitation of being conducted on simulated data due to the difficulty of obtaining such data in a real scenario. However, as an initial exploration in this direction, we show that conversational product search based on negative feedback as well as fine-grained feedback on aspect-value pairs is a promising research direction and our method of incorporating the fine-grained feedback is effective.  

There are several directions for future work. When users express their preferences on the aspects which are not asked by the system, it is important to extract the information in their conversations and combine it to refine the re-ranking. Furthermore, it is necessary to cope with users' any responses other than what the system presumes. We are also interested in studying the effect of fine-grained feedback in general question answering system after the result-level negative feedback is received. 
\begin{acks}
	This work was supported in part by the Center for Intelligent Information Retrieval and in part by NSF IIS-1715095. Any opinions, findings and conclusions or recommendations expressed in this material are those of the authors and do not necessarily reflect those of the sponsor.
\end{acks}

	\bibliographystyle{ACM-Reference-Format}
	\balance
	\bibliography{reference}
	
\end{document}